\documentclass[a4paper,10pt]{article}
\usepackage[utf8]{inputenc}
\usepackage{amsmath}
\usepackage{graphicx}
\usepackage{color}
\usepackage[margin=0.5in]{geometry}
\title{On the convergence of the Fitness-Complexity Algorithm}
\author{Emanuele Pugliese, Andrea Zaccaria, and Luciano Pietronero}
\newtheorem{guess}{Ansatz}

\begin{document}

\maketitle

\begin{abstract}
We investigate the convergence properties of an algorithm which has been 
recently proposed to measure the competitiveness of countries and the quality of 
their exported products. These quantities are called respectively Fitness $F$ 
and Complexity $Q$. The algorithm was originally based on the adjacency matrix 
$M$ of the bipartite network connecting countries with the products they export, 
but can be applied to any bipartite network. The structure of the adjacency 
matrix turns to be essential to determine which countries and products converge 
to non zero values of $F$ and $Q$. Also the speed of convergence to zero depends 
on the matrix structure. A major role is played by the shape of the ordered 
matrix and, in particular, only those matrices whose diagonal does not cross the 
empty part are guaranteed to have non zero values as outputs when the algorithm 
reaches the fixed point. We prove this result analytically for simplified 
structures of the matrix, and numerically for real cases. Finally, we propose 
some 
practical indications to take into account our results when the algorithm is 
applied.  
\end{abstract}

\section{Introduction}

Among the many theories which try to explain the features of the economic 
development of countries (see for example \cite{barro}), a recent, data driven 
and innovative approach is the one known as \emph{Economic Complexity}. This 
approach is based on the country-product matrix $M$, whose binary elements 
indicates if the given country exports a certain products or not (in particular, 
if the country shows a Revealed Comparative Advantage \cite{balassa1965trade} in 
that market). In contrast with the Ricardian specialization paradigm 
\cite{ricardo}, the $M$ matrix is, when suitably ordered, not block diagonal but 
triangular: this means that ubiquitous products are produced by all countries, 
while rare products are produced only by diversified countries. In a first 
attempt to use this information to evaluate the goodness of countries and 
products, Hidalgo and Hausmann \cite{HH} proposed an algorithm on the matrix 
$M$. Tacchella and al. \cite{Tacchella} deeply revisited the algorithm and 
proposed a new metrics to 
estimate the 
country's potential of growth, called the \emph{Fitness F}, in terms of the 
quality, or \emph{Complexity Q}, of its exported products. See \cite{plosnm} for 
a comparison between the two approaches.

\begin{minipage}{0.45\textwidth}
\begin{equation}
\label{eq:Fitness}
 \tilde{F}_c^{(n)}=\sum_p M_{cp} Q_{p}^{(n-1)}
\end{equation}
\begin{equation}
\label{eq:Complexity}
 \tilde{Q}_p^{(n)}=\frac{1}{\sum_c M_{cp} \frac{1}{F_{c}^{(n-1)}}}
\end{equation}
\end{minipage}
\begin{minipage}{0.45\textwidth}
\begin{equation}
\label{eq:FitnessNorm}
 F_c^{(n)}=\frac{ \tilde{F}_c^{(n)}}{ <\tilde{F}_c^{(n)}>_c}
\end{equation}
\begin{equation}
\label{eq:ComplexityNorm}
 Q_p^{(n)}=\frac{ \tilde{Q}_p^{(n)}}{ <\tilde{Q}_p^{(n)}>_p}
\end{equation}
\end{minipage}

where $F_c$ is the Fitness of country $c$ and $Q_p$ is the complexity of product 
$p$. All the values are determined only by the Country-Product Matrix, $M$. We 
point out that, obviously, in principle it is possible to apply this algorithm 
to any bipartite network defined by the matrix $M$, so the definitions of 
countries and products as the rows and columns of the matrix is someway 
reductive. However, since its first and main use so far is this one, in the 
following we will use the expressions country and product to denote the row and 
column elements.

In this short note we analyze the 
characteristics of $M$ required to have convergence of the algorithm to numbers 
strictly greater than zero, for both every $F_c$ and $Q_p$.
We will first show in section \ref{sec:theoretical} a class of $M$ for which it 
is possible to solve analytically the algorithm in closed form and find these 
characteristics; we therefore produce an \emph{ansatz} for the general case. We 
will then show in section \ref{sec:numerical} that the insight is numerically 
supported for any analyzed random matrix and in section \ref{sec:real} we will 
show how this is valid also for the matrices used in economic complexity. 
In the conclusions, in section \ref{sec:conclusions}, we will explain the effect 
of this result on the economic complexity field and we will give some practical 
indications for the use of the algorithm in light of our findings. 

Finally there will be an additional appendix, \ref{sec:appendixGamma}, in which 
we will try to generalize our findings to a wider class of algorithms similar to 
the one we have studied.

\section{A theoretical example}\label{sec:theoretical}

\subsection{The Matrix Class}\label{sec:matrix}
We will consider a specific class of matrices. An example of the class is 
presented in equation \ref{eq:matrix}.
The generic matrix in the class is composed of 4 blocks, one of which of zeros. 
The other three blocks will not be composed only of zeros but they will have 
some $1$s. 
The density of $1$s in the block can be in principle any value between $0$ and 
$1$, but to allow for a closed form solution each block has to have the same 
number of $1$s and $0$s in all its columns and rows.
It is interesting to note that, since the algorithm holds for reordering of rows 
and columns, a matrix is also a member of the class if it is possible to reorder 
its rows and columns in such a way to obtain a matrix in the class.

We will number the four blocks from 1 to 4 anti-clockwise starting from the top 
right corner.
In the following we will always assume that the block 4, the bottom right block, 
is the block with only zeros. Alternatively, will also refers to the block 4 as 
the \emph{external area}. 
We will also call the blocks 1 and 3 the \emph{frontier}, block 2 the 
\emph{internal area}.

In the following example of this class of matrices the four densities are, 
anti-clockwise from the top right corner, $1/2$, $1$, $1/3$, $0$.

\begin{equation}\label{eq:matrix}
\begin{array}{c}
 \quad\quad C_1 \quad\quad\quad\quad\quad C_2\\
\begin{array}{c}
\left.\begin{array}{c}
      \\
      R_2\\
      \\
      \\
      \end{array}
\right\{\\
\left.\begin{array}{c}
      \\
      R_1\\
      \\
      \end{array}
\right\{
 \end{array}
      \left(\begin{array}{ccc|cccccc}
      1&1&1&1&0&1&0&1&0\\
      1&1&1&0&1&0&1&0&1\\
      1&1&1&1&0&1&0&1&0\\
      1&1&1&0&1&0&1&0&1\\
      \hline
      0&1&0&0&0&0&0&0&0\\
      1&0&0&0&0&0&0&0&0\\
      0&0&1&0&0&0&0&0&0\\
      \end{array}
\right)
\end{array}
\end{equation}

The key variables will be the sizes of the four blocks: their height $R_1$ and 
$R_2$ and their width $C_1$ and $C_2$. 
It will be revealed instead as almost unimportant the density of $1$s in the 
non-empty blocks. 
To make the notation easier in the following we will assume that all the 
densities of the three non-empty blocks are equal to $1$: the three non empty 
blocks (1, 2 and 4) will be composed only of $1$s.
In subsection \ref{sec:density} we will explain what changes when we relax this 
assumption.

\subsection{Computations}

The symmetry of the matrix is particularly useful because we can state, for 
symmetry reasons, that all the countries in the $R_2$ upper rows have the same 
fitness, and the same is true for the countries in the $R_1$ lower rows.
This is obvious since their rows are equal or can be made equal with a simple 
rearrangement of the columns.
The same is also true for the complexity of the first $C_1$ products and the 
complexity of the last $C_2$ products.

Therefore the fitness at the iteration $n$ of the algorithm can be represented 
with just two numbers, with an obvious change of notation
\begin{equation}\label{eq:definition}
  F^{(n)}=\left(\begin{array}{c}
             a\\
             b
            \end{array}
  \right)
\end{equation}

It is easy to compute the next iteration of complexities from equation 
\ref{eq:Complexity},

\begin{equation}
  \tilde{Q}^{(n+1)}=\left(\begin{array}{c}
             \frac{ab}{R_2b+R_1a}\\
             \frac{a}{R_2}             
            \end{array}
  \right)
\end{equation}
Since we are not interested in the Complexities, to avoid unnecessary 
computations we will not normalize accordingly to \ref{eq:ComplexityNorm}.
In the next steps we will compute the fitness values and we will normalize them: 
any multiplicative factor to all the complexities would be removed in any case 
at that step.
Thanks to this property we can rewrite the Complexities in an easier way for the 
next computation, in particular as
\begin{equation}
  \tilde{Q'}^{(n+1)}=\left(\begin{array}{c}
             R_2^2b\\
             R_2^2b+R_1R_2a
            \end{array}
  \right)
\end{equation}
where $\tilde{Q'}$ is proportional to $\tilde{Q}$.

We can now compute the iteration $n+1$ for the fitness from equation 
\ref{eq:Fitness}. 
Since we are using $\tilde{Q'}$ instead of $Q$, the values obtained by equation 
\ref{eq:Fitness} will be proportional to $\tilde{F}$.
Defining these as $\tilde{F'}$, we obtain

\begin{equation}
 \tilde{F'}^{(n+1)}=\left(\begin{array}{c}
              C_1R_2^2b+C_2R_2^2b+C_2R_1R_2a\\
              C_1R_2^2b\\
            \end{array}
            \right)=\left(\begin{array}{c}
              C_1R_2^2b+C_2R_2^2b+C_2R_1^2+C_2R_1R_2-C_2R_1^2b\\
              C_1R_2^2b\\
            \end{array}
            \right)
\end{equation}
where the second step is due to imposing the normalization condition:
\begin{equation}\label{eq:normalization}
 R_1 a+R_2 b=R_1+R_2.
\end{equation}
This constraint allows us to reduce the number of variables from 2 to 1 and to 
derive a close solution. 

Finally, normalizing accordingly to equation \ref{eq:FitnessNorm} we have
\begin{equation}\label{eq:nextstep}
 F^{(n+1)}=\left(\begin{array}{c}
                  
\frac{C_1bR_2^2+C_2R_2^2b+C_2R_1^2+C_2R_1R_2-C_2R_1^2b}{
C_1bR_2^2+C_2bR_2(R_2-R_1)+C_2R_2R_1}\\
                  \frac{C_1bR_2^2}{C_1bR_2^2+C_2bR_2(R_2-R_1)+C_2R_2R_1}
                 \end{array}
                 \right).
\end{equation}

Let's focus only at the second component of the Fitness vector, which in the 
following we will call $F^{(n)}_2$, and which is referred to the fitnesses of 
the $R_1$ lowest rows. Comparing \ref{eq:definition} and \ref{eq:nextstep} we 
see that, in one algorithmic step, we had
\begin{equation}\label{eq:iteration}
 b\rightarrow 
\frac{C_1bR_2^2}{C_1bR_2^2+C_2bR_2(R_2-R_1)+C_2R_2R_1}\equiv\frac{b}{A_1b+A_2},
\end{equation}
where
\begin{equation}\label{eq:A2}
\begin{array}{rl}
  A_1=&1+\frac{C_2(R_2-R_1)}{C_1R_2}\\
  A_2=&\frac{C_2 R_1}{C_1 R_2},
\end{array}
\end{equation}
from which the closed form solution, by induction, is trivial
\begin{equation}\label{eq:closed}
F^{(n)}_2=
\frac{1}{A_1\sum_{i=0}^{n-1}A_2^i+A_2^n}.
\end{equation}
The solution for the other component can be obtained from the normalization 
condition given by Eq. \ref{eq:normalization}

\subsection{Convergence}\label{sec:shape}

How does $F^{(n)}_2$ behave when $n$ goes to infinity? It trivially depends on 
$A_2$:
\begin{itemize}
 \item if $A_2>1$, $F_2$ converges to $0$ exponentially fast while $n$ goes to 
infinity;
 \item if $A_2=1$, $F^{(n)}_2=1/(nA_1+1)$ and therefore converges to $0$ as fast 
as $n^{-1}$;
 \item if $A_2<1$, $F_2$ converges exponentially to a non zero number, 
$(1-A_2)/A_1$.
\end{itemize}

From \ref{eq:A2} we notice now that $A_2$ has also a particular geometric 
meaning:
it is the ratio of the areas of the bottom right block (block 4) and the top 
left block (block 2).
Remembering our definitions of the blocks, $A_2$ is therefore greater than $1$, 
and the second fitness component $F^{(n)}_2$ converges to $0$, if the 
\emph{external area} is bigger than the \emph{internal area}.
Or, visually, if the belly of the non-zeros area is inward.

$A_2=1$, when $A_2$ is defined through equation \ref{eq:A2}, defines the 
diagonal line, or its obvious geometric extension to rectangular matrices,
\begin{equation}
 R_1=(R_1+R_2)\frac{C_1}{C_1+C_2}.
\end{equation}

It is worth noting that if $A_2$ is close, but different, to $1$, the 
exponential behavior of the fitness can be very slow.
In particular it is possible to see that, for $A_2>1$, $F_2 \approx 
\frac{A_2-1}{A_1} e^{-(A_2-1) n}$ for $A_2$ near enough to $1$ and $n$ large 
enough. 
This define a characteristic time $n^*=1/(A_2-1)$ of the exponential that can be 
very large when $A_2$ approaches $1$.
Similarly for $A_2<1$, $F_2 \approx \frac{1-A_2}{A_1}\left(1 + e^{-n 
(1-A_2)}\right)$, again in the case of $A_2$ near enough to $1$ and $n$ large 
enough.

\subsection{Heterogeneous Density}\label{sec:density}

Interestingly the result does not change if the densities of $1$s in the blocks 
1, 2 and 3 are different than $1$, if the block 4 is still composed only of $0$s 
and the block 2 has a positive density\footnote{this condition is obviously 
needed to clearly define an internal area and an external area.}

Only the shape matters, and not the density.
In particular doing the same computations with densities different from $1$ (and 
potentially heterogeneous) we find that:
\begin{itemize}
 \item equation \ref{eq:closed} does not change,
 \item $A_2$ does not change,
 \item $A_1$ does depend on the densities of the blocks 1 and 3, but not from 
the density of block 2 if it is not $0$ (see ahead for the case in which the 
density of block 3 is zero).
\end{itemize}

The specific value of the density of $1$s and $0$s in the internal area is 
therefore completely irrelevant, while the density inside the frontier only 
determines the specific value of the convergence point (if there is convergence 
to a number greater than $0$). 

\subsubsection{Zeros outside the frontier}\label{sec:density_zeros}

The case in which the density in the internal area is $0$ is a peculiar case, 
since in this case it is not possible to really define an inside and an outside. 
In fact it is possible to rearrange the rows and columns to switch inside and 
outside. In this case it is still possible\footnote{Of course, in the case in 
which both block 2 and 4 are composed only of zeros, it is possible. to 
rearrange rows and columns to switch the blocks, i.e. there are two different 
ordered $M$, causing $A_2$ to go to $A_2^{-1}$. Therefore $A_2$ is defined but 
for a possible multiplicative inversion.} also  to define $A_2$. If $A_2$ is 
different than $1$ the fitness of one of the two sets of countries is always 
converging to zero. In the case in which $A_2$ is equal $1$ instead, any value 
of fitness is a stationary point of the algorithm and therefore the fitnesses 
produced by the algorithm will be equal to the starting conditions.

This can be said in a different way. 
In this case the blocks 1 and 3 are not connected through a product common among 
countries belonging different blocks. Therefore the algorithm in this case does 
not have any direct information to compare fitnesses and complexities of 
countries and products in different blocks.
To compare the two blocks it is therefore only possible to use their densities 
and shape. 
However the density turns out to be irrelevant, since the more products in a 
block the more is diluted their influence on the Fitnesses of the countries in 
their block: as it is possible to compute, this cancel out the additional effect 
of having more or less products.
Only the relative shape among the blocks, measured by $A_2$, is left to compare 
the two blocks. 
If $A_2$ is equal 1, and therefore also the shape does not help to compare the 
blocks, any fitness and complexity is consistent. 
If $A_2$ is different than 1, it will determine the convergence to $0$ of the 
fitnesses and complexities of one of the two blocks.

\subsection{Ansatz}\label{sec:ansatz}

The ansatz that will be then corroborated by the numerical investigation in the 
next sections is therefore simple, at least using loose definitions:

\begin{guess}If the belly of the matrix is outward all the fitnesses and 
complexities converge to numbers greater than $0$. If the belly is inward, some 
of the fitnesses will converge to $0$.
\end{guess}
 
While in the case defined in \ref{sec:matrix} the conversational definition of 
``inward belly'' - corresponding to $A_2>1$ - is unambiguous, for the general 
cases that we will investigate in the sections \ref{sec:numerical} and 
\ref{sec:real} there are many possible definitions.

Defining
\begin{itemize}
 \item \emph{ordered} matrix, as the matrix $M$ after rearranging the columns 
and rows such that all the countries are ranked accordingly to their fitnesses 
and all the products are ranked accordingly to their complexities\footnote{while 
in same cases, as we have seen and we will see, the fitness could converge to 
$0$ for many countries, for any finite number of iteration the fitness values 
will be greater than $0$. Moreover, since the convergence speed after some 
iterations is constant for each country and products, there will be a number of 
iterations such that, for any following iteration, the ranking of fitnesses and 
complexities will be constant.};
 \item \emph{diagonal} line, as the line, in the previously defined matrix, 
going from the least fit country and least complex product to the most fit 
country and most complex product; if the matrix is squared, this definition is 
the usual definition of diagonal of the matrix; if the matrix is rectangular, 
this definition is the trivial geometric extension of the previous one.
 \item the \emph{external} area of the matrix is the joint set of zeros in the 
ordered matrix including the corner corresponding to the $0$ for the lowest 
fitness and highest complexity product;
\end{itemize}
we will see that a proper definition, with relative guess, is
\begin{guess}\label{ansatz:belly}
A matrix $M$ have all the fitness and complexity different from zero if, after 
ordering said matrix, the diagonal line does not pass through the external area.
\end{guess}

It is worth to note that when a country's fitness converges to zero, from 
equation \ref{eq:Complexity}, we know that all the exported products' complexity 
will converge to zero too. At the same time, the effect on the fitnesses of 
countries which export a product whose complexity converges to zero, from 
equation \ref{eq:Fitness}, is negligible. Therefore, we can consider that, but 
for multiplicative common factors coming from equations \ref{eq:FitnessNorm} and 
 \ref{eq:ComplexityNorm}, removing from the matrix $M$ a line corresponding to a 
country converging to zero fitness and all the columns corresponding to its 
exported products, the output will be not affected.

It is therefore reasonable to make a further guess:

\begin{guess}\label{ansatz:cutpoint}
If the diagonal line does pass through the external part of a ordered matrix 
$M$, some countries and products will converge to zero. If we progressively 
remove them from the analysis, defining new ordered matrices $M$, we will have a 
convergence to finite values of fitnesses and complexities for the remaining 
countries and products for which the the new $M$ have a diagonal line not 
passing through the external area.
\end{guess}

These ansatz implicitly define the \emph{crossing country} of a matrix as the 
lowest fitness country that converges to a nonzero fitness. In other words, it 
is the first country that does not need to be removed in the removal process 
described above, defining the largest matrix such that all its countries have 
nonzero fitness.
Note that in general searching instead from the top countries does not give the 
proper result in a situation in which the diagonal cross multiple times the 
external area. If one starts the check from the top countries and keeps adding 
countries with lower fitnesses, the crossing country is not the first one met. 
In this case of multiple crossing, the crossing country is the first country met 
in the process starting from the bottom, as in ansatz \ref{ansatz:cutpoint}.

The guesses \ref{ansatz:belly} and \ref{ansatz:cutpoint} will be tested 
numerically in the sections \ref{sec:numerical} and \ref{sec:real}.

\section{A numerical investigation}\label{sec:numerical}

In this section we present the results of our numerical simulations, in which we 
study some peculiar behaviors of sample matrices which we did not treat 
analytically. In particular, by means of synthetic examples of possible $M$ 
matrices we study the cases in which more than two blocks are present, the 
importance of producing rare products and we investigate the ansatz proposed in 
the previous section. The reader will notice that the matrices investigated in 
the first two subsections correspond to the case $A_2=1$ discussed above.
\subsection{More than two blocks}
In the case of more of two blocks it is not possible to write the components of 
$F^{(n)}$ in the form given by Eq.\ref{eq:definition}, because one would have 
more than one unknown variable. However, what one can do is progressively reduce 
the given matrix to submatrices, each one made of two blocks. In the following 
example we present a block diagonal matrix: 
\[ \left( \begin{array}{cccccc}
 0&0&0&0&1&1\\
 0&0&0&0&1&1\\
 0&0&0&1&0&0\\
 1&1&1&0&0&0\\
 1&1&1&0&0&0\\   
 1&1&1&0&0&0\\ \end{array} \right)\] 
We have found that if off block diagonal elements are absent (that is, both the 
internal and the external areas are empty) all countries will remain in the 
respective initial conditions. As we expected from the two block analysis 
presented in Sec.\ref{sec:density_zeros}, this result is stable even if density 
is lower: only the relative dimensions of the blocks counts.
\subsection{The importance of oligopolies}\label{2blocks}
The presence of common products, in the very spirit of \cite{Tacchella}, does 
not add much information other than the fact that those countries which produce 
only very common products can not be very high in the fitness ranking. As a 
consequence, we can expect \textit{monopolies} of already diversified countries 
to be relevant. To investigate the consequences of having common ubiquitous 
products we take into account a synthetic situation in which three countries 
produce only very common products, two countries have a \textit{duopoly} and one 
country, which has the same diversification of these two countries, has a 
monopoly. The results are the following:
\begin{center}
 $
\bordermatrix{~ & ~ & ~ & ~ & ~ & ~ &\cr
 c&1&1&1&1&0&1\cr
 n^{-1}&1&1&1&1&1&0\cr
 n^{-1}&1&1&0&1&1&0\cr
 n^{-2}&0&1&1&0&0&0\cr
 n^{-2}&1&0&1&0&0&0\cr  
 n^{-2}&1&1&0&0&0&0\cr }
 $
\end{center}
On the left side of the matrix we show the speed of the fitness decays with the 
iteration $n$, with \textit{c} meaning convergence to a finite value of 
fitness\footnote{In this section and in the following we will not report the 
complexity decays, which follow trivially from the ones of the countries.}. For 
example, the notation $n^{\alpha}$ means that the fitness $F^{(n)}_c$ of the 
country $c$ corresponding to that row is converging towards zero with a power 
law of the number $n$ of the algorithm iterations, that is, $F^{(n)}_c \sim 
n^{\alpha}$ for a sufficiently larger $n$. As one can notice, the presence of a 
monopoly makes the first country the only one to converge to a nonzero value of 
$F$, while the advantage given by the extra products makes the second and third 
country perform better with respect to the last three countries. In any case, 
this advantage is not striking, because we are still in the case in which 
$A_2=1$, so the difference is only in the exponent of the decay. We stress that 
the 
presence of a different decay means that there exist a number $N$ so that from 
iteration $n=N$ the ranking will stay constant, and so the ranking will be given 
by the decay exponent.\\
A further, self-illustrative situation of the consequences of having common 
products is presented below.
\begin{center}
$
\bordermatrix{~ & ~ & ~ & ~ &\cr
              c & 1 & 0 & 1 & 1\cr
              n^{-1} & 1 & 1 & 1 & 0\cr
              n^{-2} & 1 & 1 & 0 & 0\cr
              n^{-3} & 1 & 0 & 0 & 0\cr}
$
\end{center}
Let us now consider the following matrix:
\begin{center}
$
\bordermatrix{~ & ~ & ~ & ~ & ~ & ~ &\cr
 c&1&0&0&0&1&1\cr
 c&0&0&0&0&1&1\cr
 n^{-0.6}&0&0&0&1&0&0\cr
 n^{-1}&0&1&1&0&0&0\cr
 n^{-1}&1&0&1&0&0&0\cr  
 n^{-1}&1&1&0&0&0&0\cr }
 $
\end{center}
here we have three blocks of nations, with one monopoly (country 3) and two 
duopolies (countries 1 and 2). On the basis of the previous considerations, we 
know that the 3x3 submatrix constituted by countries 1,2 and 3 and products 4, 5 
and 6 is degenerate, in the sense that every initial condition $F^{(0)}$ is 
stable under the iterations (also the whole matrix would be degenerate if 
country 1 would not have produced product 2). However, in this case, the 
presence of an external product changes the situation, giving an advantage to 
the duopoly, which converges to a finite value of the fitness and makes the 
monopoly tend to zero, even if with a lower exponent ($0.6$ instead of $1$). By 
means of extensive numerical simulation we have found that the specific value of 
the decay is given by the relative size of the worst block (the one which decays 
with $n^{-1}$) with respect with the total size of the externally connected 
block, that is, the sum of the sizes of the blocks which are connected by the 
external $1$. 
In other words, the decay is given by the ratio between the worst block and the 
sum of the sizes of all the blocks but the one whose decay we are computing. In 
the situation presented above, the worst block has size 3 and the converging 
block has size 2, from which we have $3/(3+2)=0.6$. To make another example we 
can imagine a different case, in which we give an advantage to the monopoly with 
respect to the block of size 3; in this case, the duopoly will decay with an 
exponent equal to $3/4$.
\\Another interesting phenomenon is evident. Let us consider the set of 
countries 4, 5 and 6. Even if the second product is owned only by countries 4 
and 5, also country 6 decays in an equally faster way. This is due to the fact 
that the last block is \emph{connected}, that is, there exist a path of products 
that connects all the countries in the block, and the absence of monopolies in 
the block. 
\subsection{Exponential decays}
Until now we have seen that different power law decays come out from, in 
general, N-polistic competitions whose symmetry is broken by the presence of 
products external with respect to the competitors. In the case in which one or 
more N-poly has a further N-poly which is not shared, the decay of the 
competitor will be \textit{exponential}. Here it is an example:
\begin{center}
$
\bordermatrix{~ & ~ & ~ & ~ & ~ &\cr
 c&0&1&0&1&1\cr
 e^{-n}&1&0&1&0&0\cr
 e^{-n}&0&1&0&0&0\cr 
 e^{-n}&1&0&0&0&0\cr }
 $
\end{center}
In this case the first country has two monopolies; country 3 one monopoly and 
countries 2 and 4 no monopolies at all. All countries tend to zero exponentially 
faster, but the first one. This is a simple example of an inward bellied matrix, 
that is, a matrix such that the diagonal crosses the external empty part. 
Because in this case the crossing country is the first one, all the other 
countries will converge to zero exponentially, as stated in the ansatz 
\ref{ansatz:cutpoint}. This phenomenon will be investigated in detail the next 
subsection.\\
\subsection{Numerical verification of the ansatz}
In Sec.\ref{sec:ansatz} we suggested a link between the convergence properties 
of countries and the shape of the belly of the ordered matrix. In this section 
we verify the given ansatz by means of simple matrices, whose behavior is 
nevertheless analogous to the one we will find in real world applications. Let 
us start with a set of 5x5 square matrices. We point out that we use square 
matrices only because the diagonal and, as a consequence, the shape of the belly 
is evident by eye. As we will see in the following, these results can be applied 
also to rectangular matrices by means of the generalized definition of diagonal 
discussed above.\\
\begin{center}
$
\bordermatrix{~ & ~ & ~ & A & ~ &\cr
 c&1&1&1&1&1\cr
 n^{-1}&1&1&1&1&0\cr
 n^{-2}&1&1&1&0&0\cr 
 n^{-3}&1&1&0&0&0\cr
 n^{-4}&1&0&0&0&0\cr
 }
 \qquad
 \bordermatrix{~ & ~ & ~ & B & ~ &\cr
 c&1&1&1&1&1\cr
 e^{-n}&1&1&1&0&0\cr
 e^{-n}&1&1&0&0&0\cr 
 e^{-n}&1&0&0&0&0\cr
 e^{-n}&1&0&0&0&0\cr
 }
 \qquad
\bordermatrix{~ & ~ & ~ & C & ~ &\cr
 c&1&1&1&1&1\cr
 c&1&1&1&1&1\cr
 c&1&1&1&1&0\cr 
 c&1&1&1&0&0\cr
 c&1&1&0&0&0\cr
 }
 \qquad
\bordermatrix{~ & ~ & ~ & D & ~ &\cr
 c&1&1&1&1&1\cr
 n^{-1}&1&1&1&1&0\cr
 n^{-1}&1&1&1&1&0\cr 
 n^{-1}&1&1&1&0&0\cr
 n^{-1}&1&1&0&0&0\cr
 }
 $
\end{center}
Matrix $A$ represents a borderline case, in which the same lines of reasoning 
adopted in Sec.\ref{2blocks} can be applied. All countries but one are 
converging to zero, preserving a well defined ranking. \\Matrix $B$ shows a 
clear inward belly. This means that, according to our ansatz, in order to know 
which countries will converge to a nonzero value of fitness we have to eliminate 
countries starting from the ones with a lowest ranking until we find a submatrix 
with an outward belly. In this case this is not possible, in the sense that the 
only fully convergent submatrix would be the trivial 1x1 matrix containing only 
country 1 and product 5. As a consequence, only the country 1 (which, in this 
case, is the trivial crossing country) and product 5 converge to a nonzero value 
of fitness and complexity, while the other countries and products decay 
exponentially.\\ On the contrary, matrix $C$ has a clear outward belly: all 
countries have products beyond the diagonal and so converge to nonzero values of 
fitness. \
An interesting situation is shown in matrix $D$, which is equal to $C$ but for 
one product. By removing product 4 from country 2 is it possible to make all 
countries but one converge to a zero value of fitness. In fact, in this case the 
diagonal crosses the external part of the matrix in correspondence with an high 
ranking country. This situation highlights the importance, in general, of a 
careful data sanitation when empirical $M$ matrices are built, not only with 
respect to low fitness countries, but also for high fitness ones. In any case, 
we point out that when one passes from matrix $C$ to $D$ the ranking is 
preserved.\\
In order to show a case in which the $M$ matrices are rectangular and the 
crossing country is nor in the top, neither in the lowest position in the 
matrix, but somewhere in the middle of it, we consider the three following 
matrices:
\begin{center}
$
\bordermatrix{~ & ~ & ~ & E & ~ &\cr
 c&1&1&1&1&1&1\cr
 c&1&1&1&1&1&0\cr
 c&1&1&1&1&0&0\cr 
 c&1&1&0&0&0&0\cr
 }
 \qquad
\bordermatrix{~ & ~ & ~ & F & ~ &\cr
 c&1&1&1&1&1&1\cr
 c&1&1&1&1&1&0\cr
 n^{-1}&1&1&1&0&0&0\cr 
 n^{-1}&1&1&0&0&0&0\cr
 }
 \qquad
 \bordermatrix{~ & ~ & ~ & G & ~ &\cr
 c&1&1&1&1&1&1\cr
 c&1&1&1&1&1&0\cr
 e^{-n}&1&1&0&0&0&0\cr 
 e^{-n}&1&1&0&0&0&0\cr
 }
 \qquad
 $
 \end{center}
 
While matrix $E$ shows full convergence of its countries, it is enough to remove 
one product to make the external area large enough to be crossed by the diagonal 
and, as a consequence, to make two countries converge to zero (matrix $F$). The 
removal of one more product turns the convergence to zero exponential with the 
number of iterations. We stress that, even if the number and the category of 
products of country 4 is the same for all the three situations, its value of 
fitness changes. However, this fact is not paradoxical as long as the values of 
the matrices are calculated in a correlated way, by considering, for example, 
the Revealed Comparative Advantage. 

\section{Real cases}\label{sec:real}
In this section we study how this variety of convergence behaviors affects real 
$M$ matrices.
Typical $M$ matrices are bigger than the ones seen in chapter 
\ref{sec:numerical} and present specific characteristics of nestedness and 
density; nevertheless, our convergence ansatz turn out to be relevant for real 
$M$ matrices.
\subsection{UN COMTRADE dataset, 1995-2010}
\begin{figure}
 \includegraphics[width=\textwidth]{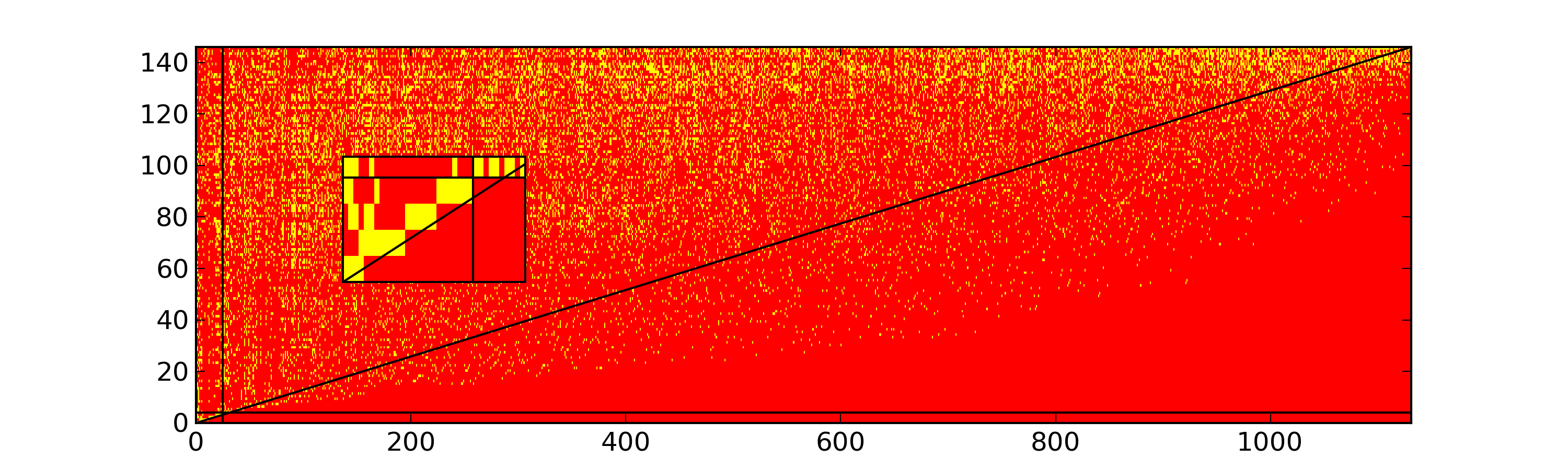}
 \caption{Ordered $M$ matrix for the BACI dataset, year 1995. The diagonal is 
consistently above the external part. However, there is an exception for the low 
fit countries, 
 zoomed in the picture inset. The horizontal and vertical lines represent the 
number of countries and products that converge to zero.\label{fig:hs2007}}
\end{figure}

The first database we consider is based on the import-export flows of products 
among countries collected by the United Nations and processed by BACI 
\cite{abbracci} and covers the years from 1995 to 2010. 
The number of products is fixed through the years and equal to 1131, classified 
accordingly to the 4 digits Harmonized System 2007, while the number of 
countries slighlty varies between 146 and 148. 
A data sanitation procedure has been performed, and the resulting values of the 
$M$ matrices are assigned by means of a threshold on the Revealed Comparative 
Advantage (RCA) as introduced by Balassa \cite{balassa1965trade}. 
In all the years the matrix is outward bellied, accordingly to the definition 
implicit in \ref{ansatz:belly}. 
In other words, the diagonal does not cross the external (i.e., empty) part of 
the matrix, and so the algorithm converges for most countries to a non zero 
value of fitness. 
While the general behavior is clear, some exceptions are there for the least fit 
countries, as it is visible in
figure \ref{fig:hs2007}. In the figure we show a pictorial representation of the 
ordered $M$ matrix for the year 1995. The yellow (red) dots mean that the 
country corresponding to the given row is (not) exporting (in the RCA sense) the 
product corresponding to the given column, that is, the matrix element is equal 
to 1 (or zero, respectively). The horizontal line corresponds to the 
\emph{crossing country} defined in \ref{sec:ansatz}, and the vertical one to the 
product with higher complexity exported by it. The part of the matrix which 
converges to nonzero values of fitness and complexity is defined by these lines 
(in particular, the top right portion of the original matrix). The dashed line 
is the diagonal of the fully convergent matrix. Obviously, after the removal 
process the matrix is outward bellied in all its parts.

It is interesting to note that the number of countries and products that need to 
be removed to have a matrix converging to non-zero values is reduced for the 
later years. This behavior is general, for all the datasets checked. The causes 
of this phenomenon will be discussed in the conclusions in section 
\ref{sec:conclusions}.

\subsection{UN COMTRADE dataset, 1963-2000}

\begin{figure}
 \includegraphics[width=\textwidth]{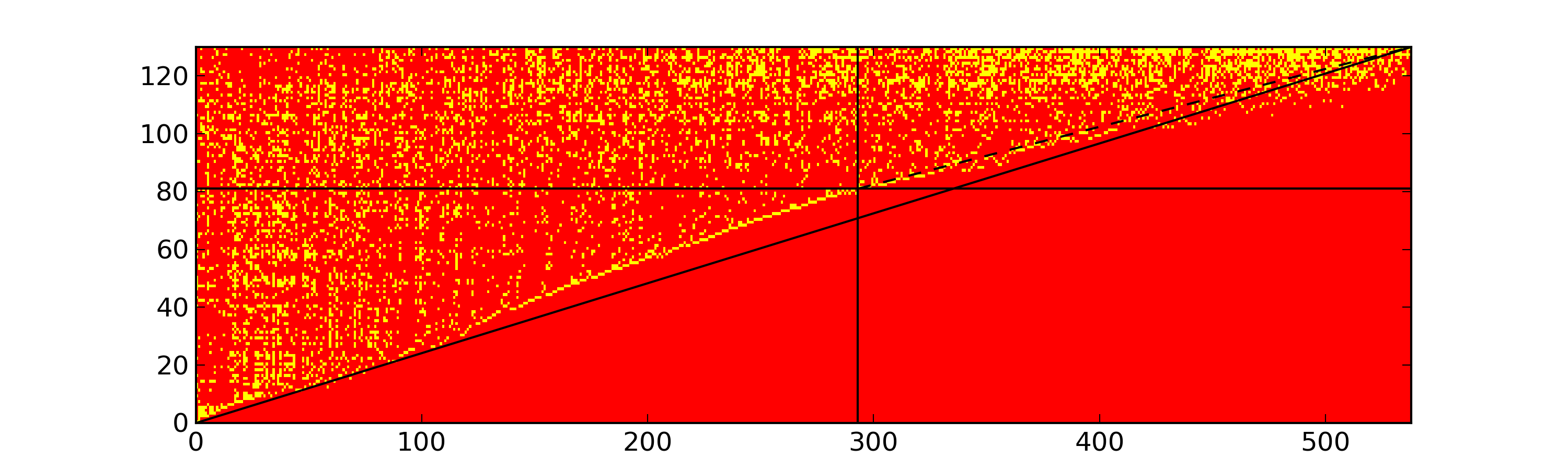}
 \caption{Ordered $M$ matrix for the 1965-2000 dataset, year 1965. 
 The matrix frontier is consistently above the diagonal line.
 The horizontal and vertical lines represent the number of countries and 
products that converges to zero.
 The dashed line shows the diagonal of the new block remaining after removing 
those countries and products: 
 the diagonal of the new $M$ matrix now does not cross the external part of the 
matrix.\label{fig:Sitc2}}
\end{figure}

A second dataset we analyze is the one collected by the United Nations and 
processed by Feenstra et al. \cite{feenstra2005world}. 
It contains the import-export flows of 538 products, classified accordingly the 
Sitc rev2 products classification, for the years 1963-2000. 
The number of countries ranges from 131 to 154. 
The dataset is particularly interesting because of the long time span of data 
presented in a consistent form. 
In this dataset the convergence ansatz proposed is more relevant, since big 
parts of it is consistently above the diagonal. 
An example, that highlight also an application of ansatz \ref{ansatz:cutpoint}, 
is presented in figure \ref{fig:Sitc2}.

The example also highlights a peculiar graphical feature of the ordered matrix. 
Indeed, for the countries and products that are converging to zero it is 
possible to observe how 
the frontier between the internal and external part is smooth and continuous, 
while a more
rough behavior can be observed for the areas converging to positive numbers.
This behavior is due to the fact that, since the fitnesses and 
complexities of the countries and products below the crossing country are 
converging exponentially to
zero, for any given finite iteration the ratio between two consecutive fitnesses 
and complexities is also
exponentially growing.
Therefore the fitnesses are mostly determined by the highest complexity product 
and, similarly, the complexities
are given by the lowest fitness country.
As a consequence there is a clear correspondence between best products and worst 
countries, determining the smooth frontier that it is 
clearly visible in figure \ref{fig:Sitc2}.
At the opposite, for the countries above the crossing country, their relative 
fitnesses converges to fixed ratios.
Therefore the fitnesses of the countries are determined by all the products not 
converging to zero, causing their ranking to be 
not uniquely determined by their best product.
A similar line of reasoning can be used for the products. 
Therefore after the crossing country there is not a clear frontier. 

A similar behavior can be observed in all the matrices that we studied. 
Even in the case of figure \ref{fig:hs2007} it is possible to recognize the 
phenomenon for the last four countries (zoomed in the inset).

\subsection{Patents}

\begin{figure}
 \includegraphics[width=\textwidth]{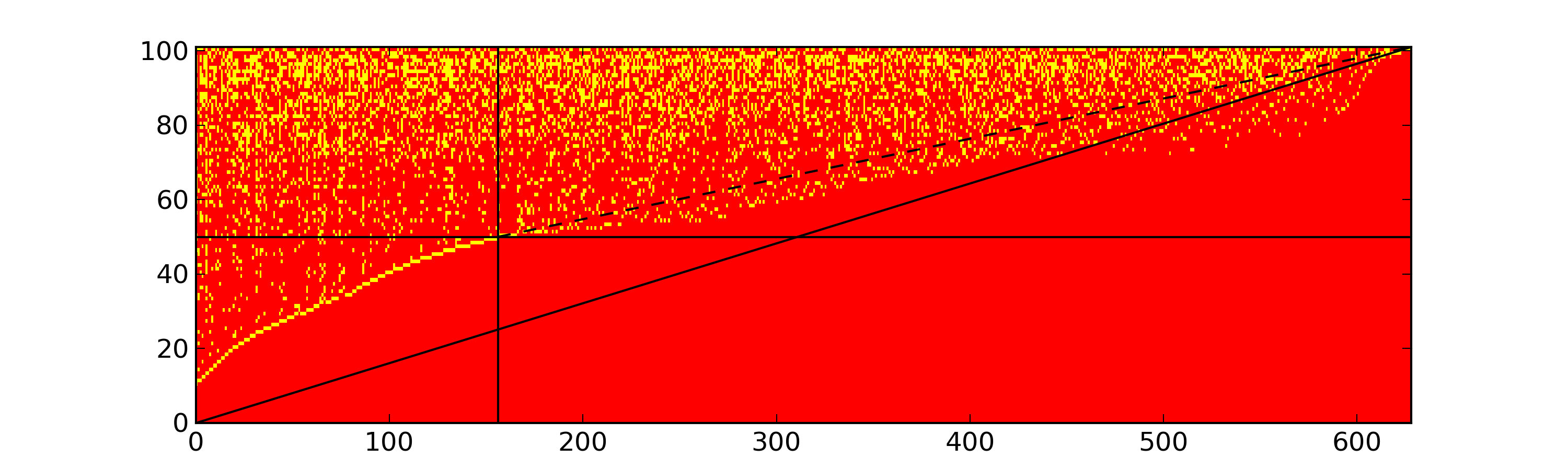}
 \caption{Ordered $M$ matrix for the patents dataset, year 2002. The external 
part is very big, so diagonal line.
 crosses it and a consistent fraction of countries converges to zero. The dashed 
line shows the diagonal of the new block remaining after removing those 
countries and products.\label{fig:patents}}
\end{figure}

The convergence to zero of a sizable set of countries is very evident for 
matrices particularly empty, like the one produce from the patents dataset. 
It is a dataset organized by OECD starting from data produced by the European 
Patent Office\footnote{For more informations, see the OECD Patent Statistics 
Manual available at http://browse.oecdbookshop.org/oecd/pdfs/free/9209021e.pdf} 
joining countries and technological sectors if a firm in a country has patents 
granted in that sector. 
It is then manipulated in the usual way, through consideration of comparative 
advantage to remove the size of the countries from the argument, 
generating a matrix that has a $1$ or $0$ if it produce more patents than 
expected in a particular technological sector. 
Since to patent a discovery in a sector the country has to be on the frontier of 
the technological progress in that sector, the matrix comes out with an inward 
belly, as it is visible in figure \ref{fig:patents}.

\begin{figure}
 \includegraphics[width=\textwidth]{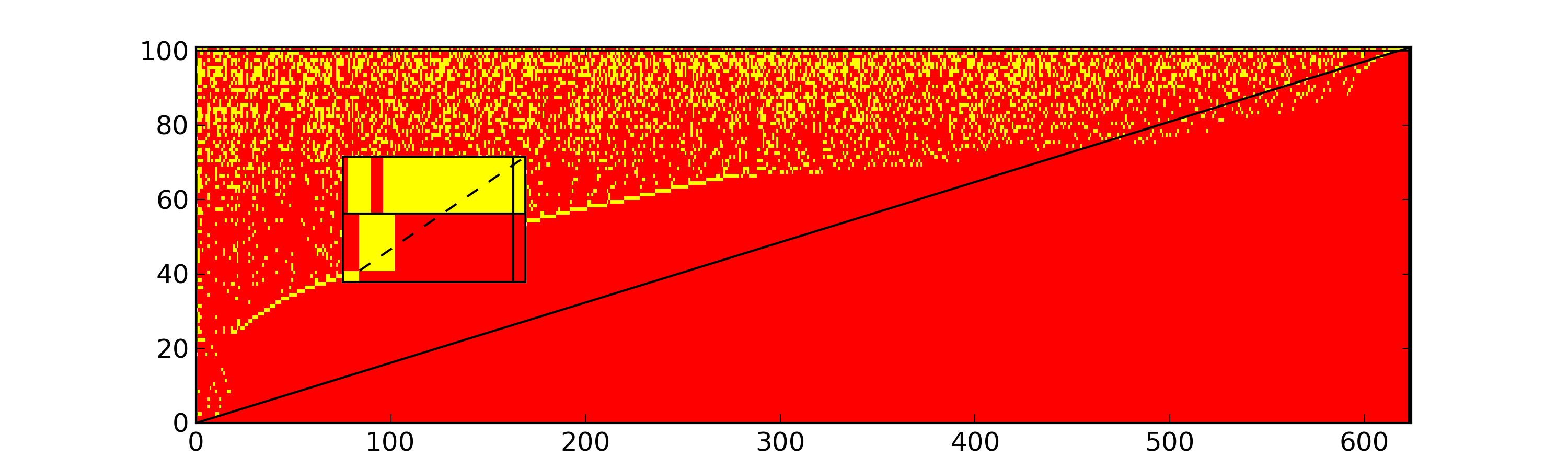}
 \caption{Ordered $M$ matrix for the patents dataset, year 2000. The diagonal 
line is consistently inside the external area. More importantly, this is true 
even for just the top two countries, as it is visible in the zoomed in the 
picture inset. As a consequence, all the countries and products converges to 
zero but the first one, as shown by the horizontal and vertical lines. The 
dashed line shows the diagonal of the block that would be left removing all 
countries and products but the first two: the diagonal would still cross the 
external part. Therefore the only remaining non-zero-fitness country is the 
first one, and the only remaining non-zero-complexity products are the ones 
produced only by it.
 \label{fig:patents2}}
\end{figure}

As for guess \ref{ansatz:cutpoint}, a possible situation leading to a global 
convergence to zero is for a single country being diversified in too many 
sectors in which no other countries is interested in. In this case we can have 
all the other countries' fitnesses converging to zero. An example is in figure 
\ref{fig:patents2}.

\section{Convergence in ranking: a practical note}

A practical application of our results is that the usual convergence conditions 
are not suitable for this algorithm. 
In particular, the condition $|F^{(n)}-F^{(n-1)}|<\epsilon$ does not imply that 
the single components of the fitness vector stop to decrease and, in general, 
the resulting fitnesses and complexities could depend on $\epsilon$.
This could be particularly relevant if one is interested in non linear 
expressions of fitness or complexities (e.g. the logarithm of the countries' 
fitness), also given the fact that convergence rates can be very slow.

In principle however, one may be interested only in the rankings. Obviously, 
different criterions can be assessed for the ranking convergence.
Indeed the problem is not trivial because of the very slow convergence of the fitness 
described in section \ref{sec:shape}. 
While the exact power law behavior 
$F_c^{(n)}=F_c^{(n_0)}\big(\frac{n}{n_0}\big)^{\alpha_c}$ described in section 
\ref{2blocks} is less likely to be found in larger real matrices, 
the exponential behavior $F_c^{(n)}=F_c^{(n_0)} e^{(n - n_0)/n^*}$ can have a 
very large characteristic time $n^*$, as noted in section \ref{sec:shape}, so 
that can be considered
approximately power law for a large number of iterations.
Therefore the ranking can change even after many iterations of the algorithm.
It is better to see the issue in a practical case.

\begin{figure}
\begin{center}
 \includegraphics[width=0.7\textwidth]{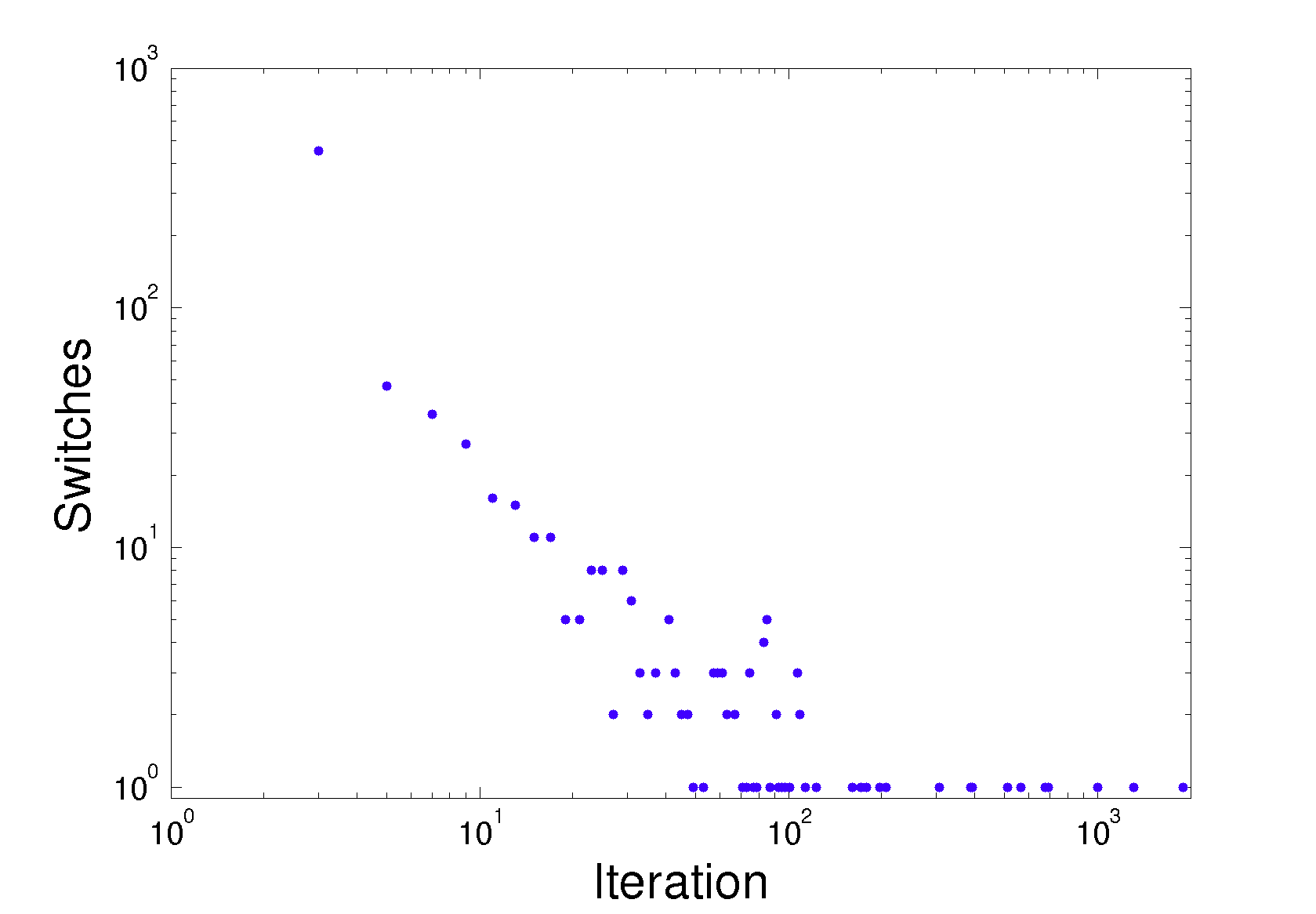}
 \caption{Number of crossings from one iteration to the next one, as a function 
of the total number of iterations.
 From the fat tail behavior shown it is clear that crossings can occour even 
after thousands of iterations.
 Data from UN COMTRADE dataset, year 1990.}\label{fig:crossing_n}
\end{center}
\end{figure}

\begin{figure}
\begin{center}
 \includegraphics[width=0.7\textwidth]{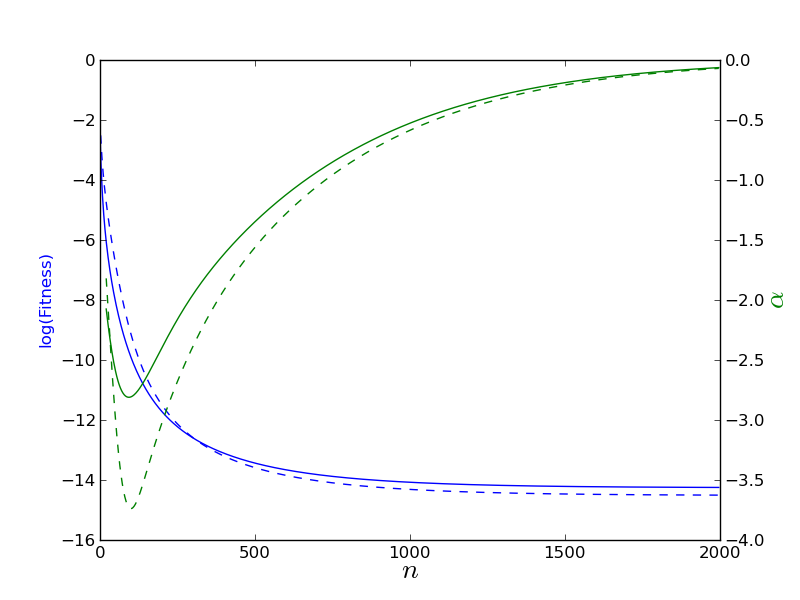}
 \caption{Comparative analysis of the behavior of the fitness and estimated 
growth rates for two sample countries near the crossing country.
 In blue the logarithm of Fitness, accordingly to the left y-axis, in green the 
growth rates.
 The two countries are identified by a solid or dashed lines.
 It is possible to see that both the growth rates and the countries' fitness 
cross after many iterations.
 In particular the ranking switches at around 500 iterations.
 It is interesting to note that the switch in rankings of the other countries 
completely changes the behavior of the two countries: indeed, until around 80 
iterations the two fitnesses seemed to exponentially converge to zero.
 However, after that point, the converging behavior emerges, as it is visible 
looking at the growth rate evolution.
 Data from UN COMTRADE dataset, year 1990.}\label{fig:fitness_alpha_evolution}
\end{center}
\end{figure}

In figure \ref{fig:crossing_n}, we show how the ranking keeps changing in a real 
case even after thousands of iteration steps. 
Indeed we can see how, in the context of a power law approximation, the 
approximated exponent $\alpha_c$ can be very small for many iterations.
Note that a value of $\alpha_c$ of $0.1$ means that 
one must wait $10^{10}$ iterations for the fitness to reduce of one order of 
magnitude. It is therefore clear that when two countries' fitness are decaying 
with
very similar $\alpha_c$, to switch ranks could take any amount of time.
We can see an example for two such countries, related to the previous real case, 
in figure \ref{fig:fitness_alpha_evolution}.

While it is impossible to assure that there will be no crossing in the next 
iterations, one can estimate a lower bound to the number of iterations still 
needed 
for the next change in ranking.
This is computed as follows:

\begin{itemize}
 \item We estimate the growth rates $\alpha_c$ for any country $c$, by the 
formula:
 \[ 
\widehat{\alpha_c}=\frac{\log(F_c^{(n)})-\log(F_c^{(n-\delta)})}{
\log(n)-\log(n-\delta)}
 \]
 where $\delta$ is a suitable delay\footnote{in the following we will use 
$\delta=2$.}.
 \item We order the countries accordingly to their fitness to have 
$F_c^{(n)}>F_{c+1}^{(n)}$.
 \item For any $c$ we estimate the crossing iteration $CI_c$ between the 
countries $c$ and $c+1$ accordingly to
 \[\label{eq:MCT}   
\widehat{CI_c}=n\bigg(\frac{F_c^{(n)}}{F_{c+1}^{(n)}}\bigg)^{1/(\widehat{\alpha_
{c+1}}-\widehat{\alpha_c})}.
 \]
 Of course the quantity $\widehat{CI_c}$ is a reasonable estimate of an actual 
possibility of a rank change only if $\widehat{CI_c}>n$, i.e. 
$\widehat{\alpha_c} > \widehat{\alpha_{c+1}}$.
 In the other case, when $\widehat{\alpha_c} < \widehat{\alpha_{c+1}}$, the two 
fitnesses are indeed diverging.
 \item Among the valid crossing iterations $\widehat{CI_c}>n$, we define the 
Minimum Crossing Iteration as 
  \begin{equation}
    MCI^{(n)}=\min_{c}(\widehat{CI_c})   
  \end{equation}
\end{itemize}

\begin{figure}
\begin{center}
 \includegraphics[width=0.7\textwidth]{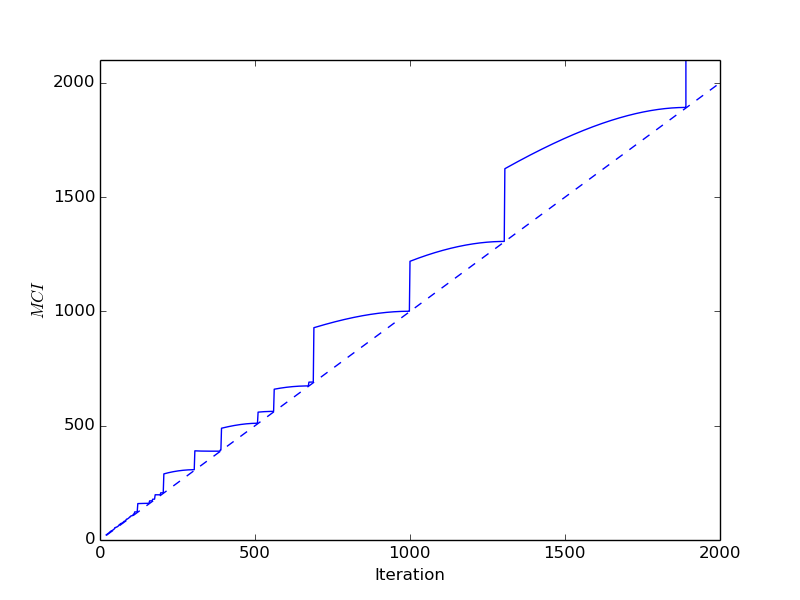}
 \caption{Evolution of the Minimum Crossing Iteration, as defined in the text, 
as a function of the iteration.
 The expected number of iterations one should wait to observe another change in 
rank appears to be a step like function jumping after each crossing (see figure 
\ref{fig:crossing_n} for a comparison).
 The increasing behavior confirms the lower bound property of the 
measure.
 After the last crossing at the iteration 1892, the next predicted change in ranking 
 is $3.7\cdot 10^6$ iterations.
 As a consequence, if I am interested only in the country ranking and I am not willing to run the algorithm for millions of iterations, 
 it is reasonable to stop the algorithm here.}\label{fig:MCI_evolution}
\end{center}
\end{figure}

The value of $MCI^{(n)}$ is therefore, for any $n$, a lower bound for the iteration 
when the next rank change will occour. Its value can be used in order to asses when to 
stop the algorithm.
This value is indeed a lower bound since the real exponential decay is always 
faster than the estimated power law. 
However, since the characteristic time of the exponential is very slow, the 
lower bound is very close to the real value.
This can be seen for example in the real case behavior of $MCI^{(n)}$ shown in figure 
\ref{fig:MCI_evolution}. 
Comparing figure \ref{fig:MCI_evolution} and figure \ref{fig:crossing_n} gives 
an insight on the algorithm. 
In particular one can precisely match the crossing times and the predicted 
minimum crossing times.
After each crossing $MCI^{(n)}$ jumps to predict the next crossing. 
Notice that the observed plateaux of $MCI^{(n)}$ show the precision of the 
power law approximation: the forecast would not change at all until the next 
crossing only if the power law behavior was exact.

As a practical note, if one is interested only in the ranking, one can set a 
threshold $MCI$ to the maximum number of iterations that he is a priori willing 
to take.
Of course it is also possible to define an equivalent measure for the ranking of 
products, that would give a different stopping time for the algorithm.

In the practical spirit of this section, it is important to stress that the 
number of iterations could get very high to achieve a satisfying value of $MCI$.
In this case it is possible that, while some countries (or products) are still 
slowly decaying, others are already well below the precision of the machine used 
for computations.
Some technical attention is therefore required to avoid that the algorithm fails 
to compute, since the Fitness-Complexity algorithm 
\ref{eq:Fitness}-\ref{eq:ComplexityNorm}, 
is ill defined for zero values of fitness.

\section{Conclusions and discussion}\label{sec:conclusions}

This paper shows that the intrinsic non linear structure of the algorithm 
proposed by \cite{Tacchella} has highly non trivial consequences on its 
convergence properties. In particular, we have linked the shape of the ordered 
matrix on which the algorithm is based to the possible presence of countries 
whose fitness tends to zero. Obviously, also all the complexities of their 
exported products will converge to zero. In synthetic but general cases it is 
possible to show analytically that the presence of a large empty part of the 
matrix makes some countries converge to zero. In particular, the condition to 
check is if the matrix diagonal crosses or not the empty part of the matrix. We 
study numerically some simple cases which confirm our results also when the 
analytical approach was not possible and in some real world cases, such as two 
different country-product databases and a country-patent database. We extend our 
results to generalized versions of the algorithm, finding similar results.
We also give a practical recipe to decide when to stop the algorithm if interested 
in a stable ranking.

A striking result of our analysis is that a large variety of situations is 
present. For example, if one considers the UN COMTRADE dataset 1995-2010, 
almost all countries have a finite fitness, while in many years of the UN COMTRADE dataset
1963-2000 more than a half of the countries converges to a zero value of fitness. On the 
other hand, in some patents datasets all the countries fitnesses but one tend to 
zero.

\bibliographystyle{apalike}       

\appendix

\section{A generalization to a wider class of 
algorithms}\label{sec:appendixGamma}

Many possible variations of the algorithm could be implemented changing 
equations \ref{eq:Fitness}-\ref{eq:ComplexityNorm} with the purpose of not 
having zero solutions.
The most obvious one is probably a variation of the harmonic mean in eq 
\ref{eq:Complexity} with a different elasticity coefficient, such as
\begin{equation}
\label{eq:ComplexityVariation}
 \tilde{Q}_p^{(n)}=\left(\sum_c M_{cp} 
\left(F_{c}^{(n-1)}\right)^{\frac{1}{\gamma}}\right)^{\gamma}.
\end{equation}
For $\gamma=-1$ equation \ref{eq:ComplexityVariation} goes back in equation 
\ref{eq:Complexity}, as in the original algorithm. 
For $\gamma=1$ equation \ref{eq:ComplexityVariation} becomes a sum, 
symmetrically to equation \ref{eq:Fitness}.
This is in strong opposition to what the algorithm is meant to describe, since 
it would mean that the more countries produce a product the higher its 
complexity.
Therefore we will limit ourselves to discuss the cases $\gamma<0$.

Computations similar to the ones in section \ref{sec:theoretical} can be done in 
this case, obtaining similar conditions.
Equation \ref{eq:iteration} is still valid, even if only in the limit $b<<a$, 
and therefore after many iterations
\footnote{the equation \ref{eq:iteration} is exactly valid for any $b$, and not 
only in the limit $b<<a$, for the case in which the density of block 2 is $0$. 
In that case, changing \ref{eq:A2} with \ref{eq:A2mod}, everything works exactly 
as in the section \ref{sec:density_zeros}}.
However the parameter $A_2$ in this case is equal to
\begin{equation}\label{eq:A2mod}
A_2=\frac{C_2}{C_1}\left(\frac{R_2}{R_1}\right)^{\gamma}.
\end{equation}

Also ansatz \ref{ansatz:belly} and \ref{ansatz:cutpoint} change accordingly.
While $A_2=1$ with $A_2$ defined in equation \ref{eq:A2} identified the diagonal 
line, when we define $A_2$ through equation \ref{eq:A2mod} we have that imposing 
$A_2=1$ we identify the curve 
\begin{equation}
 R_1=(R_1+R_2)\frac{C_2^{1/\gamma}}{C_1^{1/\gamma}+C_2^{1/\gamma}}.
\end{equation}

\begin{figure}
 \begin{center}\includegraphics[width=0.7\textwidth]{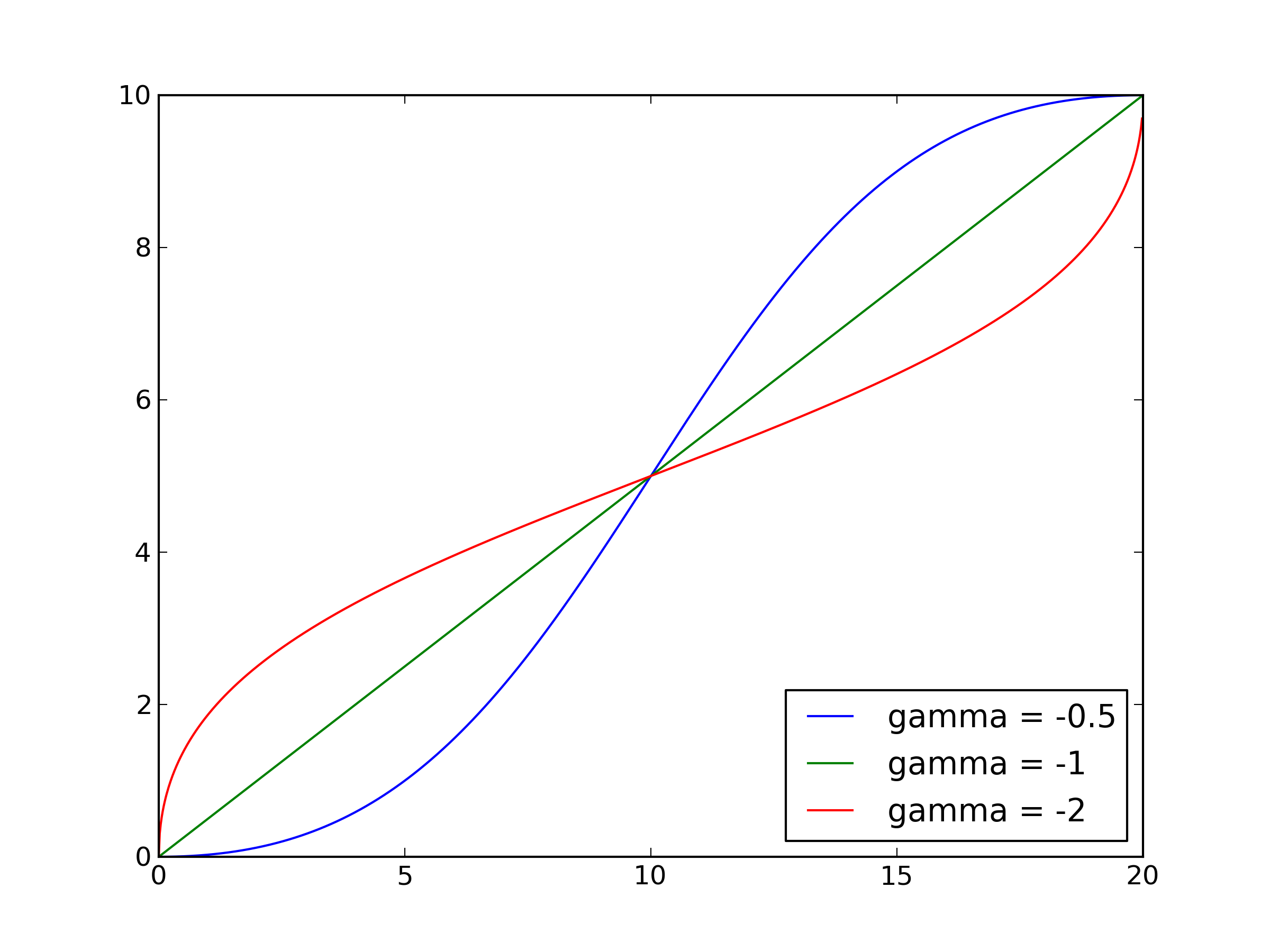}
 \caption{Line $A_2=1$ for different values of $\gamma$, for a matrix $M$ 
$10\times 20$.
 The algorithm described in equations 
\ref{eq:Fitness}-\ref{eq:ComplexityVariation} converges to all values different 
from $0$ if the frontier is all below the line.\label{fig:gammas}}
 \end{center}
\end{figure}

Examples for different values of $\gamma$ are presented in figure 
\ref{fig:gammas}. 
From the figure it is clear the trade-off of changing the value of $\gamma$ to 
avoid values of the fitness converging to $0$: to loosen the condition for high 
fitness countries, one makes it stricter for low fitness countries, and 
viceversa.
Moreover it is also clear that changing $\gamma$ does not help if the frontier 
is above the diagonal in the middle.

\end{document}